%% file: main.tex
\documentclass[11pt]{scrartcl}

\usepackage[utf8]{inputenc}

\usepackage{config}
\usepackage{titling}

\usepackage[a4paper, top=3cm, bottom=2cm, left=2.5cm, right=2.5cm, includefoot]{geometry}

\usepackage[inline]{enumitem}

\setlist[itemize]{topsep=0pt,partopsep=0pt,itemsep=0pt,parsep=0pt}
\setlist[itemize,1]{label={\small\textbullet}}
\setlist[itemize,2]{label={\tiny\textbullet}}
\setlist[itemize,3]{label=$\cdot$}
\setlist[enumerate]{topsep=0pt,partopsep=0pt,itemsep=0pt,parsep=0pt}
\setlist[enumerate,1]{label=\roman*)}
\setlist[enumerate,2]{label=\alph*)}
\setlist[enumerate,3]{label=\arabic*)}

\hypersetup{
	colorlinks=true,
	linkcolor=AO!65!black,
	citecolor=AO!65!black,
	urlcolor=AppleGreen!65!black,
	bookmarksopen=true,
	bookmarksnumbered,
	bookmarksopenlevel=2,
	bookmarksdepth=3
      }

\title{Computing the forcing spectrum of outerplanar graphs in polynomial time}
\predate{}
\date{}
\postdate{}

\preauthor{}
\DeclareRobustCommand{\authorthing}{
	\begin{center}
		Maximilian Gorsky \\
		Technische Universit\"at Berlin, Germany \\
		\href{mailto:m.gorsky@pm.me}{m.gorsky@pm.me}
  
        \medskip
        
        Fabian Kreßin \\
		Technische Universit\"at Berlin, Germany \\
		\href{mailto:fabian.kressin@tu-berlin.de}{fabian.kressin@tu-berlin.de}
\end{center}}
\author{\authorthing}
\postauthor{}

\begin{document}
\maketitle

\begin{abstract}
The forcing number of a graph with a perfect matching $M$ is the minimum number of edges in $M$ whose endpoints need to be deleted, such that the remaining graph only has a single perfect matching.
This number is of great interest in theoretical chemistry, since it conveys information about the structural properties of several interesting molecules.
On the other hand, in bipartite graphs the forcing number corresponds to the famous feedback vertex set problem in digraphs.

Determining the complexity of finding the smallest forcing number of a given planar graph is still a widely open and important question in this area, originally proposed by Afshani, Hatami, and Mahmoodian in 2004.
We take a first step towards the resolution of this question by providing an algorithm that determines the set of all possible forcing numbers of an outerplanar graph in polynomial time.
This is the first polynomial-time algorithm concerning this problem for a class of graphs of comparable or greater generality.
\end{abstract}

\section{Introduction}\label{sec:intro}
\input{introduction}

\section{Preliminaries}\label{sec:prelim}
\input{preliminaries}

\section{The matching structure of outerplanar graphs}\label{sec:outerplanar}
\input{outerplanar}

\section{Algorithm for the forcing spectrum of outerplanar graphs}\label{sec:alg}
\input{algorithms}

\section{Discussion}\label{sec:discussion}
\input{discussion}

\textbf{Acknowledgements:} This article is based on the Bachelor's thesis of the second author \cite{Kressin2023Forcing}, which was supervised by the first author.

\bibliographystyle{alphaurl}
\bibliography{literature}

\end{document}

%% file: introduction.tex

Counting the perfect matchings of a graph is one of the classic $\SharpP$-complete problems given by Valiant in \cite{Valiant1979Complexity}.
This problem has important applications in determining the structure of double-bonds in molecules and estimating their resonance energy \cite{Vukicevic2011Applications}.
In carbon-structures this is of particular interest, since these double-bounds have a substantial impact on the chemical properties of the molecule.
In light of this, Klein and Randi\'{c} \cite{RandicK1985Kekule,KleinR1987Innate} introduced the \emph{innate degree of freedom} of a molecule, represented as a graph, to try and study the impact of select subsets of the double-bonds of a molecule on its structure as a whole.
From a more mathematical perspective, Harary, Klein, and \v{Z}ivkovi\v{c} \cite{HararyKZ1991Graphical} introduced the \emph{forcing number $\Forc{G,M}$ of a perfect matching $M$} in a graph $G$, which asks for the smallest set of edges $F \subseteq M$ that have to be fixed such that $M$ is the only matching of $G$ containing $F$.

As interest in these parameters increased, the computational complexity of determining these values started gathering attention.
Dishearteningly it turns out that computing $\Forc{G,M}$ is $\NP$-complete even in bipartite graphs with maximum degree 3 \cite{AdamsMM2004Forced}.
Furthermore, in the paper that introduced the \emph{forcing spectrum} $\FSpec{G}$, which is the set of all forcing numbers $\Forc{G,M}$ for all perfect matchings $M$ of $G$, Afshani, Hatami, and Mahmoodian \cite{AfshaniHM2004Spectrum} showed that determining the minimum $\Forc{G}$ of the forcing spectrum, which was introduced as the \emph{forcing number of $G$} in \cite{HararyKZ1991Graphical}, is $\NP$-hard even for bipartite graphs with maximum degree 4.

On the positive side, Pachter and Kim \cite{PachterK1998Forcing} observed that via a well-known translation of a bipartite graph with a perfect matching into a digraph, one can compute $\Forc{G,M}$ in polynomial time using a classic result of Lucchesi and Younger \cite{LucchesiY1978Minimax} on the feedback arc set problem in digraphs and an algorithm for this problem by Gabow \cite{Gabow1995Centroids}.
However, no positive results of comparable generality are know for computing $\Forc{G}$, $\FSpec{G}$, or the \emph{maximum forcing number} $\MForc{G}$, which is the maximum of $\FSpec{G}$.
The main question in this area, which is still widely open, is therefore the following.

\begin{question}[Afshani, Hatami, and Mahmoodian \cite{AfshaniHM2004Spectrum}]\label{que:forcplanar}
    What is the complexity of determining $\Forc{G}$ for a planar graph $G$?
\end{question}

This question is already interesting if $G$ is assumed to be bipartite, considering the state of the literature on this topic.
The answer to the following question is also unknown.\footnote{In \cite{AfshaniHM2004Spectrum} this question is declared as resolved by Pachter and Kim's \cite{PachterK1998Forcing} use of the results by Lucchesi and Younger \cite{LucchesiY1978Minimax}, and Gabow \cite{Gabow1995Centroids}, but this only works for bipartite, planar graphs.}

\begin{question}\label{que:forcnonbipplanar}
    What is the complexity of determining $\Forc{G,M}$ for a non-bipartite, planar graph $G$ with a perfect matching $M$?
\end{question}

We take a first step towards the resolution of both problems by proving the following, which also directly resolves the problem of computing $\Forc{G}$ and $\MForc{G}$ for outerplanar graphs.

\begin{theorem}\label{thm:main}
    The forcing spectrum of outerplanar graphs can be computed in polynomial time.
\end{theorem}

Outerplanar graphs play a large role in theoretical chemistry.
By one estimate, at least 93,4\% of the graphs in a prominent dataset by the NCI are outerplanar \cite{HorvathRW2006Frequent}\footnote{The most current iteration of the dataset stems from 2012 and contains roughly 265.000 molecule structures, which is around 5000 more molecule structures than the dataset from 2003, which \cite{HorvathRW2006Frequent} most likely used. Thus the percentage is unlikely to have changed drastically.} and current efforts on finding forcing numbers and related parameters often focus on restricted subclasses of outerplanar graphs \cite{ZhangZL2015Forcing,ChanXN2015Linear,ZhangJ2021Continuous}.

Of particular note is the approach of studying the forcing polynomial of a given graph, introduced in \cite{ZhangZL2015Forcing}, which contains information on all perfect matchings and their forcing number.
Thus the forcing polynomial can be seen as a generalisation of the forcing spectrum.
There are several results on forcing polynomials for specific classes of graphs, though these articles generally do not mention whether the polynomials they find can be computed in polynomial time.
Here a few classes of outerplanar graphs have forcing polynomials explicit enough to make computing them feasible in polynomial time, such as zig-zag chains \cite{ZhangZL2015Forcing}, linear hexagonal chains \cite{ZhangZL2015Forcing}, $2 \times n$-grids \cite{ZhaoZ2019Forcing}, pyrene systems \cite{DengLZ2021Forcing}, and a specific class of chain-like polyomino graphs \cite{DengLW2023Forcing}.
Notably each of these classes contain at most one graph with $n$ vertices for any $n \in \N$ and all graphs considered are bipartite.

To compute the forcing number in practice, the most prominent method is exhaustive enumeration (for example see Figure 2 in \cite{VukicevicR2005Kekule}), with an integer linear programming formulation for the problem being proposed recently \cite{LiuMYW2022Computing}.
Since solving integer linear programs is famously an $\NP$-complete problem \cite{Karp1972Reducibility}, this does not affect the current state of \Cref{que:forcplanar} and \Cref{que:forcnonbipplanar}.

We are not aware of any results which explicitly concern computing the forcing spectrum of a graph.
Furthermore, positive computational results for the forcing number are rare and tend to be tied to characterisations of specific graph classes (see \cite{HansenZ1994Bonds} and \cite{WangYZW2008Forcing} for examples).
For a helpful survey of the area surrounding the forcing number see \cite{CheC2011Forcing}.

Our approach is somewhat orthogonal to the currently popular methods for the forcing number, deriving more from classic matching theory, as it is presented in \cite{LovaszP1986Matching}, and algorithmic approaches from structural graph theory.
To facilitate our efforts, we first show in \Cref{sec:outerplanar} that computing the forcing number of a graph $G$ can always be reduced to computing the forcing number of its \emph{cover graph} $\Cov{G}$, which is the spanning subgraph of $G$ that only contains those edges which are found in a perfect matching of $G$.

\begin{lemma}\label{lem:forccovergraph}
    For any graph $G$ with a perfect matching $M$, we have $\Forc{G,M} = \Forc{\Cov{G},M}$ and $\FSpec{G} = \FSpec{\Cov{G}}$.
\end{lemma}

Since the cover graph is computable in polynomial time \cite{CarvalhoC2005VE}, this is a sensible reduction for this problem that can be applied at little cost.
We then use this to study the matching structure of outerplanar graphs, in particular with respect to the tight cut decomposition introduced by \Lovasz \cite{Lovasz1987Matching}.
A connected graph is called \emph{matching covered} if all of its edges are contained in perfect matchings of the graph.
This leads us to the following conclusion.

\begin{lemma}\label{lem:outerplanarmeansbip}
    Any matching covered, outerplanar graph is 2-connected and bipartite.
\end{lemma}

Since the tight cut decomposition can be computed in polynomial time \cite{Lovasz1987Matching}, we are able to present a dynamic programming algorithm on this decomposition to compute the forcing spectrum in \Cref{sec:alg}.
We choose the tight cut decomposition for this, since we need to control matchings and tight cuts identify parts of the graph where the structure of perfect matchings is very restricted.
However, although dynamic programming is a standard technique in algorithmic applications of structural graph theory, the tight cut decomposition is not exactly the most suited candidate for this approach.
On top of this, since we want to compute the forcing spectrum, we have to somehow consider each of the possibly exponentially many perfect matchings of the outerplanar graph in our polynomial-time computation.
For this reason, the proof of correctness for our algorithm is rather involved.

We close the article with a discussion in \Cref{sec:discussion} on whether our approach can be generalised to other graph classes and the hurdles to extending our results to the anti-forcing number, which is an edge-deletion version of the forcing number.

%% file: preliminaries.tex

Our notation for graphs largely orients itself on \cite{BondyM1976Graph}.
All graphs considered in this thesis are simple and thus contain neither parallel edges nor loops.
By $C_4$ we denote the cycle of length 4.

If the vertex set $V(G)$ of a graph $G$ can be partitioned into two sets $U$ and $W$ such that any edge $uw \in E(G)$ has one endpoint in $U$ and the other in $W$, we call $G$ \emph{bipartite}.

For a vertex set $X \subseteq V(G)$, we denote by $G - X$ the graph $\InducedG{G}{V(G) \setminus X}$ and for an edge set $F \subseteq E(G)$, the graph $G - F$ is defined as $(V(G), E(G) \setminus F)$.
If $F$ is a set of edges, we write $V(F)$ to denote the set of endpoints of the edges of $F$.

\subsection{Outerplanar graphs}\label{subsec:outerplanar}
A \emph{planar} graph is a graph with a drawing in the plane in which no two edges cross each other in anything besides their endpoints.
By deleting all points and curves contained in the drawing of a planar graph we split the plane into several connected regions which are called \emph{faces} of the drawing.
The \emph{boundary} of a face is the subgraph consisting of all vertices and edges whose corresponding curves and points of the planar drawing are completely contained in the closure of that face.
If there exists a planar drawing of a graph in which all vertices are found in the boundary of the same face, we call such a graph \emph{outerplanar}, we call this face the \emph{outer face}, and the corresponding boundary is the \emph{outer boundary}.


The structure of outerplanar graphs is well understood and there exist many easy to prove facts which will be helpful to us.
In particular we need the following two statements.

\begin{lemma}[folklore]\label{lem:2degenerate}
    All outerplanar graphs contain a vertex of degree at most 2.
\end{lemma}

We call a graph $G$ \emph{Hamiltonian} if there exists a cycle $H$ that contains all vertices of $G$.
Such a cycle is called \emph{Hamiltonian}.
As a corollary to Whitney's result \cite{Whitney1932Nonseparable} that for any 2-connected planar graph all faces are bounded by cycles, we also get the following well-known fact.

\begin{corollary}\label{lem:hamiltonicity}
    The outer boundary of a 2-connected, outerplanar graph is a Hamiltonian cycle.
\end{corollary}

\subsection{Matching theory}\label{subsec:matchingtheory}
A \emph{matching} in a graph is a set of mutually disjoint edges and such a matching is called \emph{perfect} if every vertex of the graph is contained in an edge of the matching.
By $\Perf{G}$ we denote the set of perfect matchings of $G$.
If every edge in a connected graph $G$ with at least four vertices is contained in some perfect matching, we call $G$ \emph{matching covered}.

\begin{theorem}[Plummer \cite{Plummer1980Nextendable}]\label{thm:2con}
    Any matching covered graph is 2-connected.
\end{theorem}

Given a graph $G$ with a perfect matching, we construct the \emph{cover graph $\Cov{G}$} by removing all edges of $G$ that are not found in any perfect matching of $G$.
By definition the cover graph of $G$ is matching covered.
The cover graph of any graph can be computed in $\mathcal{O}(mn)$-time via an algorithm by Carvalho and Cheriyan \cite{CarvalhoC2005VE}.
For a perfect matching $M \in \Perf{G}$, a cycle $C \subseteq G$ is called $M$-alternating if $E(C) \cap M$ is a perfect matching of $C$.

Let $G$ be a graph, let $X \subseteq V(G)$ be a vertex set, and let $\CutG{G}{X}$ be the set of edges in $G$ which have exactly one endpoint in $X$.
We call $\CutG{G}{X}$ the \emph{cut around $X$ (in $G$)} and if $G$ is clear from the context, we simply write $\Cut{X}$.
Given $X$ and $G$, we let $\Compl{X}$ be $V(G) \setminus X$.

In a matching covered graph $G$, a cut $\Cut{X}$ is called \emph{tight} if $|\Cut{X} \cap M| = 1$ for all $M \in \Perf{G}$.
A tight cut $\Cut{X}$ is called \emph{trivial} if $X$ or $\Compl{X}$ contains exactly one vertex.
We note in particular that for any tight cut $\Cut{X}$ both $X$ and $\Compl{X}$ contain an odd number of vertices.
Two tight cuts $\Cut{X}$ and $\Cut{Y}$ are called \emph{laminar}, if $A \subseteq B$ or $B \subseteq A$ holds for some $A \in \{ X, \Compl{X} \}$ and $B \in \{ Y, \Compl{Y} \}$.

\begin{definition}[Tight cut contraction]
    Let $G$ be a matching covered graph with a tight cut $\Cut{X}$.
    The two \emph{tight cut contractions} $\ContractXinGtoV{X}{G}{c}$ and $\ContractXinGtoV{\Compl{X}}{G}{c}$ of $G$ at $\Cut{X}$ are constructed by contracting the vertices in $X$, respectively in $\Compl{X}$, into the vertex $c$ and deleting all loops and parallel edges that result from this.
    We generally assume that $c$ is chosen such that $c \not\in V(G)$.
    
    Equivalently, $\ContractXinGtoV{X}{G}{c}$ can be constructed by deleting $X$, introducing $c$, and adding all edges between $c$ and vertices in $\Compl{X}$ that are incident to an edge in $\Cut{X}$.
    We note that for any trivial tight cut one of the two tight cut contractions will be the graph itself and the other is $K_2$.
    This serves to justify that we call these tight cuts trivial.
\end{definition}

Tight cut contractions preserve a myriad of properties, a few of which are listed here.

\begin{lemma}\label{lem:propertypreserved}
    Let $G$ be a matching covered graph with a tight cut.
    Then
    \begin{enumerate}
        \item any tight cut contraction of $G$ is also matching covered,

        \item if $G$ is outerplanar, any tight cut contraction of $G$ is outerplanar, and

        \item $G$ is bipartite if and only if all of its tight cut contractions are bipartite.
    \end{enumerate}
\end{lemma}

Of particular importance to structural matching theory are those matching covered graphs on at least four vertices that only have trivial tight cuts.
If such a graph is bipartite, we call it a \emph{brace} and if it is non-bipartite, it is instead called a \emph{brick}.
In this article we will only need to discuss braces and we note that $C_4$ is the unique smallest brace.

This now allows us to introduce the \emph{tight cut decomposition} due to \Lovasz \cite{Lovasz1987Matching}.
Given a matching covered graph $G$, we can repeatedly take tight cut contractions at non-trivial tight cuts until we are left with a list of graphs on at least four vertices which do not possess non-trivial tight cuts.
The graphs in this list are the \emph{braces} and respectively \emph{bricks of $G$}.

In \cite{Lovasz1987Matching}, \Lovasz argued that this decomposition can be computed in polynomial time and showed that the list of bricks and braces is independent of the choice of the tight cuts at which we contract.
In particular, this means that a tight cut decomposition can always be performed by finding a maximal family of laminar tight cuts in $G$ and contracting along these cuts.

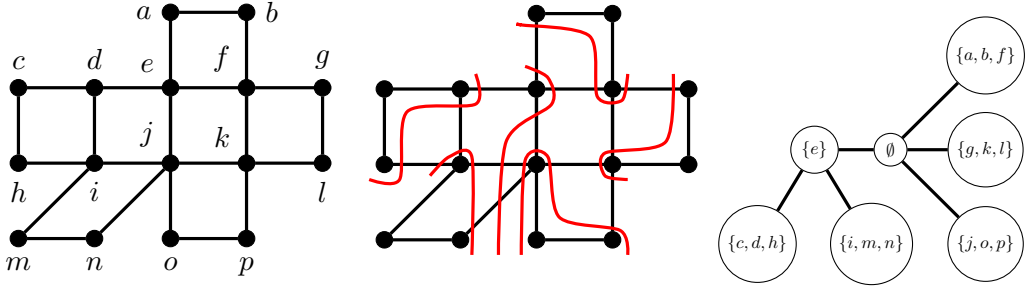
\begin{figure}[h!]
	\begin{center}
		\input{graphs/tightcutexample} \
        \raisebox{7.75pt}{\input{graphs/tightcutslaminar}} \ 
        \input{graphs/tightcutdecomp}
	\end{center}
	\caption{A matching covered, outerplanar graph on the left, with a laminar set of tight cuts marked in red in the middle, and a corresponding tight cut decomposition following \Cref{def:tightcutdecomp} on the right.
    The tight cuts assigned to the edges are given implicitly, since deleting any edge in the tree gives a partition of the vertices of the graph into two sets.}
	\label{fig:decomp}
\end{figure}

The tight cut decomposition is usually given in the informal way found in the previous paragraph.
As we want to perform a dynamic programming algorithm on this decomposition, we need a more formal definition.
An example of such a decomposition is given in \Cref{fig:decomp}.

Given a tree $T$ and an edge $uv \in E(T)$, we denote by $T_u$ and $T_v$ the two components of $T - uv$, where $T_u$ is the component containing $u$ and $T_v$ is the component containing $v$.

\begin{definition}[Tight cut decomposition]\label{def:tightcutdecomp}
    Let $\FFF$ be a maximal family of laminar tight cuts in a matching covered graph $G$, and let $T$ be a tree on $|\FFF| + 1$ vertices.
    Then $(T, \beta , \pi )$ is a \emph{tight cut decomposition} if $\pi \colon E(T) \rightarrow \FFF$ is a bijection such that
    \begin{enumerate}
        \item $\beta \colon V(T) \rightarrow 2^{V(G)}$ partitions $V(G)$, where we explicitly allow empty sets, and\label{item:partition}

        \item for all $e = uv \in E(T)$ with $\pi(e) = \Cut{X}$, we have $\bigcup_{w \in V(T_u)} \beta(w) = X$ and $\bigcup_{w \in V(T_v)} \beta(w) = \Compl{X}$ or vice versa.\label{item:cuts}
    \end{enumerate}
\end{definition}

We note that the \emph{bags}, meaning the sets $\beta(v)$ for $v \in V(T)$, can have arbitrary size and the same is true for the tight cuts we associate with the edges of $T$.
Controlling these two parts of the decomposition will ultimately be what leads us to a polynomial time algorithm.



The decomposition also allows us to recover the bricks and braces of the graph as follows.

\begin{definition}[Bricks and braces from a tight cut decomposition]
    Let $G$ be a matching covered graph with a tight cut decomposition $(T, \beta , \pi )$.
    For any $v \in V(T)$, we let $br(v)$ be the result of contracting all sets $\bigcup_{w \in V(T_u)} \beta(w)$ in $G$ for every edge $uv \in \CutG{T}{\{v\}}$.
\end{definition}

To prove that this definition is sensible, we will need a small technical lemma on the translation of tight cuts across tight cut contractions.

\begin{lemma}\label{lem:tightcuttranslation}
    Let $G$ be a matching covered graph with a tight cut $\CutG{G}{X}$ and let $G' = \ContractXinGtoV{X}{G}{c}$ with a tight cut $\CutG{G'}{Y}$.
    Then $F \coloneqq (\CutG{G'}{Y} \cap E(G)) \cup \{ uv \in E(G) \setminus E(G') \mid uc \in E(G') \}$ is a tight cut in $G$ that is laminar to $\CutG{G}{X}$.
\end{lemma}
\begin{proof}
    For $\CutG{G'}{Y}$ we observe that either $Y \subseteq V(G)$ or $\Compl{Y} \subseteq V(G)$.
    W.l.o.g.\ we will suppose that $Y \subseteq V(G)$.
    From this we can deduce that $F$ is in fact a cut $\CutG{G}{Y}$ in $G$.
    Furthermore, since $X$ is contracted in $c$ in $G'$, we thus have either $c \in Y$ or $c \in \Compl{Y}$ and therefore $\CutG{G}{X}$ and $\CutG{G}{Y}$ are guaranteed to be laminar.

    Suppose that $\CutG{G}{Y}$ is not tight.
    We first observe that since $|Y|$ is odd, for every perfect matching $M \in \Perf{G}$ there must exist at least one edge in $M \cap \CutG{G}{Y}$ and in fact $|M \cap \CutG{G}{Y}|$ must be odd.
    Let $M$ be a perfect matching of $G$ such that $M \cap \CutG{G}{Y}$ contains at least three distinct edges $e,f,g$.
    We can translate $M$ into a perfect matching $M' = (M \cap E(G')) \cup \{ uc \mid h = uv \in M \text{ and } |h \cap X| = 1 \}$, where the latter set in the construction contains exactly one edge, as $\CutG{G}{X}$ is tight.

    If $|\{ e,f,g \} \cap M'| \geq 2$, this would contradict $\CutG{G'}{Y}$ being a tight cut in $G'$.
    We may therefore suppose w.l.o.g.\ that we have $e,f \not\in M'$.
    But this means that $e,f \not\in M \cap E(G')$, which implies that $|e \cap X| = 1 = |f \cap X|$ and thus there exist two distinct edges $e', f' \in \{ uc \mid h = uv \in M \text{ and } |h \cap X| = 1 \} \subseteq M'$, which contradicts the construction.
    Thus $\CutG{G}{Y}$ must be tight.
\end{proof}

\begin{lemma}\label{lem:bracerecovery}
    Let $G$ be a matching covered graph with a tight cut decomposition $(T, \beta , \pi )$.
    Then for all $v \in V(T)$ the graph $br(v)$ has no non-trivial tight cuts.
\end{lemma}
\begin{proof}
    Suppose towards a contradiction that $br(v)$ has a non-trivial tight cut.
    Then according to \Cref{lem:tightcuttranslation} this induces a cut $\Cut{X}$ in $G$ that is laminar to all other tight cuts in $\FFF$, where $\FFF$ is the maximal family of laminar tight cuts that forms the image of $\pi$, which contradicts the maximality of $\FFF$.
\end{proof}

\subsection{The forcing number}\label{subsec:forcing}
\begin{definition}[The forcing number]\label{def:force}
    Given a graph $G$ and some $M \in \Perf{G}$, a \emph{forcing set $F \subseteq E(G)$ for $M$} is a set of edges such that $M \setminus F$ is the unique perfect matching of $G - V(F)$.
    The \emph{forcing number of $M$ in $G$}, denoted as $\Forc{G,M}$ is the minimum size of a forcing set $F \subseteq M$ for $M$ and the \emph{forcing-spectrum} $\FSpec{G}$ is then simply defined as $\{ \Forc{G,M} \mid M \in \Perf{G} \}$.
    
    The \emph{(minimum) forcing number} $\Forc{G}$ is $\min (\FSpec{G})$ and the \emph{maximum forcing number} $\MForc{G}$ is $\max (\FSpec{G})$.
\end{definition}

Within our algorithm we will want to guess parts of a smallest forcing set for a given perfect matching.
This however means that some parts of the matching are not forced by this set of edges and we will need terminology to discuss how these edges behave.
Given a set $F \subseteq M$ in a graph $G$ with a perfect matching, we say that an edge $e \in E(G) \setminus F$ is \emph{forced in (by $F$)} if $e$ is contained in every perfect matching of $G - V(F)$ and $e$ is \emph{forced out (by $F$)} if it is contained in no perfect matching of $G - V(F)$.
An edge that is forced in or out is also simply called \emph{forced}.
Any edge that is not forced by $F$ will be called \emph{open (with respect to $F$)}.
In particular this means that for any open edge $e$ there exist two distinct perfect matchings $M_1, M_2 \in \Perf{G - V(F)}$ with $e \in M_1$ and $e \not\in M_2$.

%% file: graphs/tightcutexample.tex
			\begin{tikzpicture}

            \node (U2) at (2,3) [draw, circle, scale=0.6, fill, label=west:{$a$}] {};
            \node (U3) at (3,3) [draw, circle, scale=0.6, fill, label=east:{$b$}] {};

            \node (V0) at (0,2) [draw, circle, scale=0.6, fill, label=north:{$c$}] {};
            \node (V1) at (1,2) [draw, circle, scale=0.6, fill, label=north:{$d$}] {};
            \node (V2) at (2,2) [draw, circle, scale=0.6, fill, label=north west:{$e$}] {};
            \node (V3) at (3,2) [draw, circle, scale=0.6, fill, label=north west:{$f$}] {};
            \node (V4) at (4,2) [draw, circle, scale=0.6, fill, label=north:{$g$}] {};

            \node (W0) at (0,1) [draw, circle, scale=0.6, fill, label=south:{$h$}] {};
            \node (W1) at (1,1) [draw, circle, scale=0.6, fill, label=south:{$i$}] {};
            \node (W2) at (2,1) [draw, circle, scale=0.6, fill, label=north west:{$j$}] {};
            \node (W3) at (3,1) [draw, circle, scale=0.6, fill, label=north west:{$k$}] {};
            \node (W4) at (4,1) [draw, circle, scale=0.6, fill, label=south:{$l$}] {};

            \node (A0) at (0,0) [draw, circle, scale=0.6, fill, label=south:{$m$}] {};
            \node (A1) at (1,0) [draw, circle, scale=0.6, fill, label=south:{$n$}] {};
            \node (A2) at (2,0) [draw, circle, scale=0.6, fill, label=south:{$o$}] {};
            \node (A3) at (3,0) [draw, circle, scale=0.6, fill, label=south:{$p$}] {};

            \foreach\i in {0,...,3}
                {
                    \pgfmathsetmacro\iplus{\i+1}
                    \path (V\i) edge[very thick] (V\iplus);
                    \path (W\i) edge[very thick] (W\iplus);
                    \path (V\i) edge[very thick] (W\i);
                }
                \path (V4) edge[very thick] (W4);
            \foreach\i in {2,3}
                {
                    \path (U\i) edge[very thick] (V\i);
                    \path (V\i) edge[very thick] (A\i);
                }
            \path
                (U2) edge[very thick] (U3)
                (A2) edge[very thick] (A3)
                (A0) edge[very thick] (W1)
                (A1) edge[very thick] (W2)
                (A0) edge[very thick] (A1)
            ;
			\end{tikzpicture}

%% file: graphs/tightcutslaminar.tex
			\begin{tikzpicture}

            \node (U2) at (2,3) [draw, circle, scale=0.6, fill] {};
            \node (U3) at (3,3) [draw, circle, scale=0.6, fill] {};

            \node (V0) at (0,2) [draw, circle, scale=0.6, fill] {};
            \node (V1) at (1,2) [draw, circle, scale=0.6, fill] {};
            \node (V2) at (2,2) [draw, circle, scale=0.6, fill] {};
            \node (V3) at (3,2) [draw, circle, scale=0.6, fill] {};
            \node (V4) at (4,2) [draw, circle, scale=0.6, fill] {};

            \node (W0) at (0,1) [draw, circle, scale=0.6, fill] {};
            \node (W1) at (1,1) [draw, circle, scale=0.6, fill] {};
            \node (W2) at (2,1) [draw, circle, scale=0.6, fill] {};
            \node (W3) at (3,1) [draw, circle, scale=0.6, fill] {};
            \node (W4) at (4,1) [draw, circle, scale=0.6, fill] {};

            \node (A0) at (0,0) [draw, circle, scale=0.6, fill] {};
            \node (A1) at (1,0) [draw, circle, scale=0.6, fill] {};
            \node (A2) at (2,0) [draw, circle, scale=0.6, fill] {};
            \node (A3) at (3,0) [draw, circle, scale=0.6, fill] {};

            \foreach\i in {0,...,3}
                {
                    \pgfmathsetmacro\iplus{\i+1}
                    \path (V\i) edge[very thick] (V\iplus);
                    \path (W\i) edge[very thick] (W\iplus);
                    \path (V\i) edge[very thick] (W\i);
                }
                \path (V4) edge[very thick] (W4);
            \foreach\i in {2,3}
                {
                    \path (U\i) edge[very thick] (V\i);
                    \path (V\i) edge[very thick] (A\i);
                }
            \path
                (U2) edge[very thick] (U3)
                (A2) edge[very thick] (A3)
                (A0) edge[very thick] (W1)
                (A1) edge[very thick] (W2)
                (A0) edge[very thick] (A1)
            ;

            \node (C1) at (-0.2,0.8) [draw=none, label={}] {};
            \node (C2) at (1.2,2.2) [draw=none, label={}] {};
            \node (Con1) at (0.2,0.8) [draw=none, label={}] {};
            \node (Con2) at (0.3,1.7) [draw=none, label={}] {};
            \node (Con3) at (1.2,1.8) [draw=none, label={}] {};
            \draw[very thick, red] plot [smooth] coordinates {(C1) (Con1) (Con2) (Con3) (C2)};

            \node (C1) at (0.6,0.85) [draw=none, label={}] {};
            \node (C2) at (1.15,-0.2) [draw=none, label={}] {};
            \node (Con1) at (0.85,1.1) [draw=none, label={}] {};
            \node (Con2) at (1.15,1.1) [draw=none, label={}] {};
            \draw[very thick, red] plot [smooth] coordinates {(C1) (Con1) (Con2) (C2)};

            \node (C1) at (1.85,2.3) [draw=none, label={}] {};
            \node (C2) at (1.5,-0.2) [draw=none, label={}] {};
            \node (Con1) at (2.15,2.15) [draw=none, label={}] {};
            \node (Con2) at (2.2,1.8) [draw=none, label={}] {};
            \node (Con3) at (1.6,1.3) [draw=none, label={}] {};
            \draw[very thick, red] plot [smooth] coordinates {(C1) (Con1) (Con2) (Con3) (C2)};

            \node (C1) at (1.8,-0.2) [draw=none, label={}] {};
            \node (C2) at (3.2,-0.2) [draw=none, label={}] {};
            \node (Con1) at (1.85,1.05) [draw=none, label={}] {};
            \node (Con2) at (2.15,1.1) [draw=none, label={}] {};
            \node (Con3) at (2.3,0.3) [draw=none, label={}] {};
            \node (Con4) at (3.1,0.125) [draw=none, label={}] {};
            \draw[very thick, red] plot [smooth] coordinates {(C1) (Con1) (Con2) (Con3) (Con4) (C2)};

            \node (C1) at (1.8,2.85) [draw=none, label={}] {};
            \node (C2) at (3.2,2.2) [draw=none, label={}] {};
            \node (Con1) at (1.8,2.85) [draw=none, label={}] {};
            \node (Con2) at (2.7,2.7) [draw=none, label={}] {};
            \node (Con3) at (2.8,1.9) [draw=none, label={}] {};
            \node (Con4) at (3.125,1.85) [draw=none, label={}] {};
            \draw[very thick, red] plot [smooth] coordinates {(C1) (Con1) (Con2) (Con3) (Con4) (C2)};

            \node (C1) at (3.8,2.2) [draw=none, label={}] {};
            \node (C2) at (3.2,0.8) [draw=none, label={}] {};
            \node (Con1) at (3.7,1.3) [draw=none, label={}] {};
            \node (Con2) at (2.9,1.15) [draw=none, label={}] {};
            \node (Con3) at (2.9,0.85) [draw=none, label={}] {};
            \draw[very thick, red] plot [smooth] coordinates {(C1) (Con1) (Con2) (Con3) (C2)};
            
            \end{tikzpicture}

%% file: graphs/tightcutdecomp.tex
			\begin{tikzpicture}
   
            \node (T1) at (3.25,2.5) [draw, circle, scale=0.6] {$ \{ a,b,f \} $};
            \node (T2) at (1,1.25) [draw, circle, scale=0.6] {$ \{ e \} $};
            \node (T3) at (2,1.25) [draw, circle, scale=0.6] {$ \emptyset $};
            \node (T4) at (3.25,1.25) [draw, circle, scale=0.6] {$ \{ g,k,l \} $};
            \node (T5) at (0.25,0) [draw, circle, scale=0.6] {$ \{ c,d,h \} $};
            \node (T6) at (1.75,0) [draw, circle, scale=0.6] {$ \{ i,m,n \} $};
            \node (T7) at (3.25,0) [draw, circle, scale=0.6] {$ \{ j,o,p \} $};

            \path
                (T1) edge[very thick] (T3)
                (T2) edge[very thick] (T3)
                (T2) edge[very thick] (T5)
                (T2) edge[very thick] (T6)
                (T3) edge[very thick] (T4)
                (T3) edge[very thick] (T7)
            ;
			\end{tikzpicture}

%% file: outerplanar.tex

We start with a general observation that shows that when trying to determine the forcing number for a given graph it suffices to consider its cover graph.

\begin{lemma}
    If $G$ is a graph with a perfect matching $M$, then $\Forc{G,M} = \Forc{\Cov{G},M}$.
\end{lemma}
\begin{proof}
    Let $F \subseteq M$ be a forcing set for $M$ in $\Cov{G}$ and suppose that $F$ is not a forcing set for $M$ in $G$.
    Then there exists a perfect matching $M'$ in $G - V(F)$ that contains an edge $e \in E(G) \setminus E(\Cov{G})$.
    However this implies that there exists a perfect matching $N \in \Perf{G}$ with $e \in N$ but $e \not\in E(\Cov{G})$, a contradiction.

    For the other direction we need to merely observe that $\Cov{G} \subseteq G$ and thus any forcing set for $M$ in $G$ must be forcing for $M$ in $\Cov{G}$.
\end{proof}

\Cref{lem:forccovergraph} is an immediate corollary of this result and we can thus concentrate on the subclass of outerplanar graphs that is matching covered.
Thanks to \Cref{thm:2con}, we also know that all of these graphs are 2-connected.
We can further narrow down the structure of these graphs.

\begin{lemma}\label{lem:onlyc4braces}
    Let $G$ be a matching covered, outerplanar graph.
    Then $G$ only has braces and all braces of $G$ are $C_4$.
\end{lemma}
\begin{proof}
    The fact that $C_4$ is a brace can easily be checked by confirming that all of its tight cuts are trivial.
    To find a tight cut in $G$, we use \Cref{lem:2degenerate} and choose a vertex $v \in V(G)$ that has degree at most 2.
    Since $G$ is matching covered it is also 2-connected, thanks to \Cref{thm:2con}, and thus $v$ has exactly two distinct neighbours $u$ and $w$.
    
    We now observe that $\Cut{\{u,v,w\}}$ is a tight cut, since $v$ must be matched either to $u$ or to $w$ in each perfect matching of $G$.
    The tight cut contraction associated with $\Cut{\{u,v,w\}}$ that contains $v$ must be $C_4$.
    The other tight cut contraction is assured to be outerplanar according to \Cref{lem:propertypreserved}.
    We can therefore repeat this procedure until we have decomposed $G$ into a list of copies of $C_4$.
\end{proof}

From this and \Cref{lem:propertypreserved} we can easily derive \Cref{lem:outerplanarmeansbip}.
We also get the following nice result.

\begin{corollary}
    There are no outerplanar bricks and the only outerplanar brace is $C_4$.
\end{corollary}

Since \Cref{lem:outerplanarmeansbip} tells us that the relevant graphs are bipartite, this suggests an easy extension of the approach of Pachter and Kim in \cite{PachterK1998Forcing} in which we first compute all perfect matchings of an outerplanar graph, for each of them translate the graph to a digraph, and solve the directed feedback arc set problem on these digraphs.
As long as there are only polynomially many perfect matchings, in terms of the number of vertices of the input graph, this would provide an easy brute-force algorithm to compute the forcing spectrum of outerplanar graphs, provided we can find all of these matchings in polynomial time.
However, it is not hard to find an outerplanar graph with exponentially many perfect matchings.

\begin{figure}[h!]
	\begin{center}
		\input{graphs/ladder}
	\end{center}
	\caption{A ladder of length 16 with the edges of a perfect matching, chosen as in the proof of \Cref{lem:exponentialpms}, marked in red.}
	\label{fig:ladder}
\end{figure}
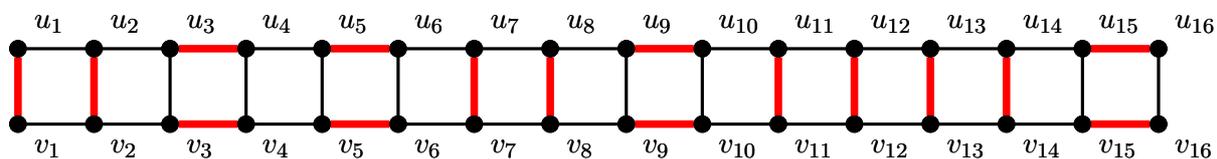

Let $k \in \N$ be an integer and let $P$ and $Q$ be two paths of length $k$, with the vertex sets $V(P) = \{ v_1, \ldots , v_k \} $ and $V(Q) = \{ u_1, \ldots , u_k \}$, and edge sets $E(P) = \{ v_iv_{i+1} \mid i \in [ k-1 ] \}$ and $E(Q) = \{ u_iu_{i+1} \mid i \in [ k-1 ] \}$.
We construct a \emph{ladder of length $k$} by taking the union of $P$ and $Q$ and introducing the edges $\{ v_iu_i \mid i \in [ k ] \}$ (see \Cref{fig:ladder} for an example).
It is easy to verify that for any positive $k$ the ladder of length $k$ is outerplanar, 2-connected, and matching covered.

\begin{lemma}\label{lem:exponentialpms}
    For any positive, even $k \in \N$, the ladder of length $k$ has at least $2^{\nicefrac{k}{2}}$ distinct perfect matchings.
\end{lemma}
\begin{proof}
    We note that since $k$ is even, there exists a packing of disjoint facial cycles of length 4 that cover the entire graph $G$.
    Each of these cycles has two perfect matchings and we can choose these matchings independent of the other cycles in the packing.
    Thus, if the packing contains $\ell$ cycles, there are at least $2^\ell$ distinct perfect matchings for the entire graph $G$.
    It is easy to verify that the ladder of length $k$ can be covered with a packing of exactly $\nicefrac{k}{2}$ such facial cycles, completing the proof of our statement.
    The only $k$ for which the number of matchings meets the stated lower bound is $k = 2$.
\end{proof}

This serves to justify the effort we will put into further analysing the structure of these graphs and our algorithm.
For the remainder of this section we study the structure of tight cuts in outerplanar graphs.
First, let us introduce an order on the edges of a tight cut in this setting.

\begin{definition}[Order $<_X$ on a tight cut $\Cut{X}$]\label{def:tightcutorder}
    Let $G$ be a matching covered, outerplanar graph with a tight cut $\Cut{X}$ and let $H$ be the Hamiltonian cycle in $G$ that bounds the outer face.
    There exist exactly two distinct edges $e,f \in \Cut{X} \cap E(H)$ and $H - \{ e,f \}$ consists of two paths $P_X$ and $P_{\Compl{X}}$, with $P_X \subseteq \InducedG{G}{X}$ and $P_{\Compl{X}} \subseteq \InducedG{G}{\Compl{X}}$.
    
    Any edge in $\Cut{X}$ has one endpoint in $V(P_X)$ and the other in $V(P_{\Compl{X}})$.
    For further use, we orient both of these paths from their endpoint in $e$ towards their endpoint in $f$.
    This structure now induces the following total order $<_X$ over the edges in $\Cut{X}$:
    For any two edges $g,h \in \Cut{X}$ we set $g <_X h$ if the endpoint of $g$ in $V(P_X)$ occurs before the endpoint of $h$ in $V(P_X)$ or the endpoint of $g$ in $V(P_{\Compl{X}})$ occurs before the endpoint of $h$ in $V(P_{\Compl{X}})$.
    
    We note that at least one condition must hold, since $G$ does not have parallel edges, and if both conditions contradict each other, this implies the existence of two edges that cross in the drawing.
\end{definition}

The existence of this order may seem like a somewhat strange technical observation, but this order allows us to prove that edges in a tight cut $\Cut{X}$ that are open, after the removal of some partial forcing set $F$ on one side of $\Cut{X}$, must form an interval in $\Cut{X}$ with respect to the order $<_X$.
Thus we can reduce the number of potential sets of open edges that need to be forced later on in our algorithm from exponentially many down to polynomially many.

\begin{lemma}\label{lem:interval}
    Let $G$ be a matching covered, outerplanar graph with a tight cut $\Cut{X}$ and a perfect matching $M$.
    Further, let $F \subseteq M \cap E(\InducedG{G}{X})$.
    There do not exist three distinct edges $f,e,g \in \Cut{X}$ with $f <_x e <_x g$, such that $e$ is forced, and $f$ and $g$ are open.
\end{lemma}
\begin{proof}
    Towards a contradiction, suppose that this triple of edges does exist.
    If $e$ is forced in, then since $\Cut{X}$ is a tight cut, $f$ and $g$ cannot be open.
    Thus we may suppose that $e$ is forced out.

    Let $N$ be a perfect matching in $G$ with $F \cup \{ f \} \subseteq N$, which must exists, since $f$ is open.
    Similarly, there must exist a matching $N' \in \Perf{G}$ with $F \cup \{ g \} \subseteq N'$.
    Since $\Cut{X}$ is tight and $f,g \in \Cut{X}$, there exists a cycle $C_f$ that contains $f$ and is alternating for both $N$ and $N'$.
    For analogous reasons, there must also exists a cycle $C_g$ that contains $g$ and is alternating for $N$ and $N'$.
    If $C_f \neq C_g$ then each cycle must use at least two edges of $\Cut{X}$ and thus $N$ or $N'$ uses at least two edges of $\Cut{X}$, contradicting the fact that $\Cut{X}$ is tight.
    Thus we have $C_f = C_g$ and furthermore, all edges of $C_f$ have to be open with respect to $F$, since $C_f$ is alternating for $N$ and $N'$.
    In particular this means that $F \cap E(C_f) = \emptyset$.

    As $G$ is outerplanar and the drawing of $C_f$ in the outerplanar drawing of $G$ is a closed curve, we also know that both endpoints of $e$ are found in $V(C_f)$.
    We note that, according to \Cref{lem:outerplanarmeansbip}, $G$ is bipartite and thus $C_f \cup e$ contains two distinct even cycles that both use $e$.
    One of these two cycles is $N$-alternating and the other is $N'$-alternating.
    Either of them allows us to find a perfect matching $N''$ that contains both $F$ and $e$, contradicting the fact that $e$ is forced out.
\end{proof}

\begin{lemma}
    Let $G$ be a matching covered, outerplanar graph with a tight cut $\Cut{X}$ and a perfect matching $M$.
    Further, let $F \subseteq M \cap E(\InducedG{G}{X})$.
    The number of different sets of edges from $\Cut{X}$ that are not forced out by $F$ is polynomial in $|\Cut{X}|$. 
\end{lemma}
\begin{proof}
    Of course there exist exactly $p \coloneqq |\Cut{X}|$ different edges that can be forced in.
    Thus we only need to count the number of edge sets within $\Cut{X}$ that consist of open edges with respect to $F$.
    As an immediate consequence of \Cref{lem:interval}, any such set of edges must be an interval of edges with respect to $<_X$, adding up to at most $\nicefrac{p(p+1)}{2} \in \mathcal{O}(p^2)$ possible sets.
\end{proof}

%% file: graphs/ladder.tex
			\begin{tikzpicture}

            \foreach\i in {1,...,16}
                {
                    \node (V\i) at (\i,0) [draw, circle, scale=0.6, fill, label=south east:{$v_{\i}$}] {};
                    \node (U\i) at (\i,1) [draw, circle, scale=0.6, fill, label=north east:{$u_{\i}$}] {};
                }

            \foreach\i in {1,...,15}
                {
                    \pgfmathsetmacro\iplus{\i+1}
                    \path (V\i) edge[very thick] (U\i);
                    \path (V\i) edge[very thick] (V\iplus);
                    \path (U\i) edge[very thick] (U\iplus);
                }
            \path (V16) edge[very thick] (U16);

            \path
                (V1) edge[red, line width=1mm] (U1)
                (V2) edge[red, line width=1mm] (U2)
                (V3) edge[red, line width=1mm] (V4)
                (U3) edge[red, line width=1mm] (U4)
                (V5) edge[red, line width=1mm] (V6)
                (U5) edge[red, line width=1mm] (U6)
                (V7) edge[red, line width=1mm] (U7)
                (V8) edge[red, line width=1mm] (U8)
                (V9) edge[red, line width=1mm] (V10)
                (U9) edge[red, line width=1mm] (U10)
                (V11) edge[red, line width=1mm] (U11)
                (V12) edge[red, line width=1mm] (U12)
                (V13) edge[red, line width=1mm] (U13)
                (V14) edge[red, line width=1mm] (U14)
                (V15) edge[red, line width=1mm] (V16)
                (U15) edge[red, line width=1mm] (U16)
            ;

            \foreach\i in {1,...,16}
                {
                    \node (A\i) at (\i,0) [draw, circle, scale=0.6, fill, label=south east:{$v_{\i}$}] {};
                    \node (B\i) at (\i,1) [draw, circle, scale=0.6, fill, label=north east:{$u_{\i}$}] {};
                }
			\end{tikzpicture}

%% file: algorithms.tex

We will need a few more bits of notation to present the algorithm.
Given a tree $T$, we root this tree at some vertex $root(T)$ and this induces an order over the vertices of $T$, such that for each $v \in V(T)$ there exists a set of vertices $children(v)$, which consists of the neighbours of $v$ that are not found on the unique path from $root(T)$ to $v$, and there exists a unique $parent(v)$, for all $v \in V(T) \setminus \{ root(T) \}$.
Note that for the leaves of $T$, the set $children(v)$ is empty.

Given a tight cut $\Cut{X}$ in a matching covered, outerplanar graph $G$, we use the order $<_X$ to associate with each edge $e \in \Cut{X}$ an integer from $[ |\Cut{X}| ]$ that corresponds with its place in the order $<_X$ and denote this value as $index(<_X,e)$.
We will also need the indicator function $\mathbbm{1}$, where $\mathbbm{1}(x,y) \coloneqq 1$ for two integers $x,y \in \N$ if $x=y$ and otherwise $\mathbbm{1}(x,y) \coloneqq 0$.


\subsection{Statement of the algorithm}\label{subsec:algostate}

Though dynamic programming is a standard technique for the construction of algorithms, we will nonetheless thoroughly discuss our approach, since to our knowledge this is the first algorithm performing dynamic programming directly on a tight cut decomposition and tight cut decompositions were not exactly intended for this purpose.
A particular problem is that we will have to frequently switch contexts between the brace corresponding to a particular bag, the original graph in its entirety, and a partial contraction of the original graph, to prove the correctness of our method.

\begin{algorithm}
\caption{Finding the forcing-spectrum of a matching covered, outerplanar graph.}\label{alg:forcing}
\begin{algorithmic}[1]
\Procedure{ForcingSpectrum}{matching covered, outerplanar graph $G = (V, E)$}

\State Find a tight cut decomposition $(T,\beta,\pi)$ of $G$ and designate some leaf of $T$ as $root(T)$.
\State Let $n = |V|$, let $m = |E|$, and let $r = root(T)$.

\ForAll{ $v \in V(T) \setminus \{ r \}$ }
        \State Let $X_v \subseteq V(G)$ be the unique set such that $\Cut{X_v} = \pi(\{ v, parent(v) \})$ and $\beta(r) \subseteq \Compl{X_v}$.
\EndFor
\State Let $c$ be the unique child of $r$ in $T$.
\State Let $w \in \beta(r)$ be the unique vertex that is not incident to any edge of $\Cut{X_c}$. 
\State Let $X_r = V(G) \setminus \{ w \}$.
\State Set $A[\Cut{X_v},e_v,(a,b),x] = 0$ for all $v \in V(T)$, $e_v \in \! \Cut{X_v}$, $(a,b) \in \! [|\Cut{X_v}|]^2$, and $x \in [\frac{n}{2}]$.

\While{there exists an unmarked $v \in V(T)$ without unmarked children}
    
    \State Let $\Cut{Y_i} = \pi( vw_i )$ for all $w_i \in children(v)$ and let $k = |children(v)|$.
    \State Let $S = \{ \pi(e) \mid e \in \CutG{T}{\{ v \}} \} \cup \bigcup\limits_{w \in \beta(v)} \{ \CutG{G}{\{w\}} \}$.
    \State Let $A, B, C, D \in S$ be pairwise distinct with $A \cap C = B \cap D = \emptyset$.
    \State Let $S' = \{ \{ e_1, e_2 \} \mid ( e_1 \in A \cap B $ and $ e_2 \in C \cap D )$ or $( e_1 \in A \cap D $ and $ e_2 \in B \cap C ) \} $.
    
    \ForAll{matchings $M \in S'$}
    
        \State Let $\{e_v\} = \Cut{X_v} \cap M$ and let $\{f_v\} = M \setminus \{ e_v \}$.

        \If{ $k=0$ }
            \State $A[\Cut{X_v}, e_v, (1, |\Cut{X_v}|), 0] \leftarrow 1$
            \State \textbf{continue}
        \EndIf
        
        \State Let $\{e_i\} = \Cut{Y_i} \cap M$ for all $i \in [k]$.
        \State Let $T_i = \{ ((a,b),x) \mid A[\Cut{Y_i}, e_i, (a,b), x] = 1 \}$ for all $i \in [k]$.
             
        \ForAll{tuples $(t_1,...,t_k) \in T_1 \times ... \times T_k$}

                \State Let $(a_i, b_i) = t_i[1]$ and let $Z_i = \{ e \mid a_i \leq index(<_{Y_i},e) \leq b_i \}$ for all $i \in [k]$.
                \State Let $W = \{ M' \in S' \mid $ for every $i \in [k] $ we have $|M' \cap Z_i| = 1 \}$.
                \State Let $W_e = \{ M' \in W \mid e_v \in M' \}$.
                \State Let $W_f = \{ M' \in W \mid f_v \in M' \}$.
                
                \If{ $|W_e| = 1$ }
                    \State Let $F = \emptyset$ and let $Z = \bigcup\limits_{M' \in W} (M' \cap \Cut{X_v})$.
                \Else
                    \State Let $F = \{ f_v \}$ and let $Z = \bigcup\limits_{M' \in W_f} (M' \cap \Cut{X_v})$.
                \EndIf

                \State Let $a$ and $b$ be the indices of the first, respectively last, edge of $Z$ w.r.t.\ $<_X$.
                \State $A[\Cut{X_v}, e_v, (a,b), |F| + \sum_{i=1}^k t_i[2]] \leftarrow 1$
        \EndFor
    \EndFor
    \State Mark $v$ as visited.
\EndWhile

\State \Return $\{ x + 1 - \mathbbm{1}(a,b) \mid  \textrm{There exist } {x \in [\frac{n}{2}]}, \ {(a,b) \in [ | \Cut{X_r} | ]^2}, \textrm{ and } {e \in \Cut{X_r}}$
        \Statex \hspace{4.75cm} with ${A[\Cut{X_r}, e, (a,b), x] = 1} . \}$

\EndProcedure
\end{algorithmic}
\end{algorithm}

We start by giving an abstract explanation of \Cref{alg:forcing}.

\textbf{Lines 1--8:}
Our input is a matching covered, outerplanar graph and thus the tight cut decomposition $(T, \beta, \pi)$ has the property that for all $v \in V(T)$ the graph $br(v)$ is a $C_4$, according to \Cref{lem:onlyc4braces} and \Cref{lem:bracerecovery}.
In particular this means that $T$ has maximum degree 4.
We root $T$ in a leaf $r$.
Notably, for every leaf $v \in V(T)$, we know that $|\beta(v)| = 3$.
When defining the tight cuts $\Cut{X_v}$ for all $v \in V(T) \setminus \{ r \}$, we choose $X_v$ such that $X_v \cap \beta(r) = \emptyset$.
For the root we choose $X_r$ in a fashion that makes it easier to compute the forcing spectrum at the end of the algorithm.

\textbf{Line 9:}
If we are looking at an entry $A[\Cut{X_v}, e_v, (a,b), x]$ of the table $A$, this intends to convey the following information:
\begin{itemize}
    \item We are considering a partial forcing set for the part of the graph corresponding to $X_v$.

    \item Our intention is to force $e_v$ or at least contribute towards forcing this edge.

    \item The partial forcing set leaves an interval of edges from $\Cut{X_v}$ open, beginning on the edge with index $a$ and ending on the edge with index $b$ with respect to $<_{X_v}$.

    \item The size of the partial forcing set we used is $x$.
\end{itemize}
If the entry $A[\Cut{X_v}, e_v, (a,b), x]$ is set to 1, then this conveys that we did actually find a partial forcing set with the effects and size described above.

\textbf{Lines 10--14:}
The while-loop simply traverses $T$ in a bottom-up fashion.
At this point of the algorithm, we use the structure of $br(v)$ to figure out the behaviour of the set of edges corresponding to the tight cuts and vertices that make up the structure of $br(v)$ in the original graph $G$.
The set $S$ contains all tight cuts that are contracted into distinct vertices of $br(v)$ and the edges incident to all vertices $w \in V(G)$ that are also found in $br(v)$ and are thus not the result of a tight cut contraction.
Note that the tight cut $\Cut{X_v}$ is always found in $S$.
Since $br(v)$ is the $C_4$, this means that $S$ contains exactly four pairwise distinct edge sets $A,B,C,D$, whose names can be chosen such that $A \cap C = B \cap D = \emptyset$, since there are two pairs of vertices in $br(v)$ that occupy the same colour classes.
As $A,B,C,D$ therefore stand in for the four vertices of $C_4$, we collect in $S'$ all matchings in $G$ that correspond to either of the two possible perfect matchings of $C_4$.
Note that while there are only two perfect matchings of $C_4$, there are many pairs of edges in $G$ that ultimately get contracted into the edges of $br(v)$ that represent these two matchings.

\textbf{Lines 15--19:}
For each matching in $S'$ there exists one edge $e_v$ that lies in $\Cut{X_v}$, since $\Cut{X_v} \in S$, and the other edge $f_v$ lies outside of $\Cut{X_v}$.
Should $v$ be a leaf of $T$ and $v \ne r$, we can force the edge $f_v$ later on in our algorithm.
Thus we set the entry in $A$ to 1 that says that $\emptyset$ is the partial forcing set we use here.

\textbf{Lines 20--23}:
For each child associated with a tight cut $\Cut{Y_i}$, we check which edge $e_i \in M$ is found in the matching.
Note that two tight cuts can both use $e_i$.
We then check the table to see what the partial forcing set associated with $e_i$ in the tight cut $\Cut{Y_i}$ has cost us and what other edges of $\Cut{Y_i}$ remain open.
Here it has to be mentioned that there can be more than one possible range $(a,b)$ and associated cost $x$.
In particular, there may be two possible intervals of open edges in $\Cut{Y_i}$ that do not contain each other.
For all combinations of possible costs and ranges of edges of the different tight cuts associated with the children of $v$ in $T$, we then check how we can force out the open edges other than $e_i$ in the tight cuts $\Cut{Y_i}$ and what the resulting costs and new open edges in $\Cut{X_v}$ are.

\textbf{Lines 24--26}:
In $W$ we collect all matchings in $S'$, which consist of edges in $G$, that use edges from the sets of open edges that are left with our current selection of partial forcing sets.
Within this set, we then filter out two types of matchings.
First, $W_e$ contains all those matchings that use $e_v$, which is the edge which we are supposed to force in $\Cut{X_v}$.
This set represents the set of matchings in $S'$ we narrow ourselves down to if we take $e_v$ into our partial forcing set.
Second, $W_f$ consists of those matchings that use $f_v$, representing those matchings we get if we take $f_v$ into our partial forcing set.
We note that these sets are non-empty, since the matching $M$ (from line 15) is contained in all of them.

\textbf{Lines 27--28}:
Should $W_e$ contain exactly one matching, then, as pointed out above, this matching must be $M$ and therefore we can force the edge $f_v$ by taking $e_v$ into our partial forcing set.
Since we defer paying for $e_v$ until we deal with the parent of $v$ in the algorithm, we save that the new set of forcing edges $F$ is empty and save the new set of open edges in $Z$, which consists of all edges from $\Cut{X_v}$ that are found in matchings from $W$, since we did not take any new edge into our partial forcing set. 

\textbf{Lines 29--30}:
Alternatively, if $W_e$ contains more than one matching, we must take $f_v$ into our partial forcing set.
We update $F$ accordingly, and since we added $f_v$ into our partial forcing set, we can determine $Z$ by only considering the matchings in $W_f$.

\textbf{Lines 31--34}:
At the ends of our loops, we add the size of $F$ to the cost of all the partial forcing sets we incurred along the way and encode $Z$ via the indices given to it by $<_{X_v}$.
This is then saved in $A$ and once we have dealt with all matchings in $S'$, we mark $v$ as visited.
Finally, we return all values $x$ from the table $A$ for all edges $e$ of $\Cut{X_r}$, which is the set of incident edges of the vertex in $\beta(r)$ that is not incident to the tight cut edges belonging to the child of $r$, as chosen in the beginning.
If for some reason the very last edge $e$ is not already forced, which is indicated by the fact that $a = b$, we must also take $e$ into our forcing set, which causes the final cost to increase by exactly one for this particular case.

\subsection{Correctness and runtime of the algorithm}\label{subsec:algoproof}

We are now ready to prove the correctness of our algorithm.
There are two core claims within this proof that we need to verify:
First, we need to show that the forcing set our algorithm finds is minimal.
This is complicated by the fact that there can be more than one minimal forcing set and the algorithm of course finds a very specific type of forcing set.
Second, we of course need to show that what we find is in fact a forcing set and that we accurately assess its size, as we proceed through the tight cut decomposition.

\begin{theorem}\label{thm:correctnessalgo}
    \Cref{alg:forcing} returns the forcing spectrum of the matching covered, outerplanar graph $G$ that is given as its input.
\end{theorem}
\begin{proof}
    Let $(T, \beta, \pi)$ be the tight cut decomposition found for $G$ and let $r = root(T)$ be a leaf, as chosen in the algorithm.
    We define $\Cut{X_v}$ as in lines 4--8 of the algorithm.

    Let $v \in V(T)$ be an arbitrary vertex and let $\Cut{X_v}$ be the tight cut associated with it.
    For the remainder of the proof, we will need to relate certain edge sets of $br(v)$, to edge sets of $G$, and edge sets of $\ContractXinGtoV{\Compl{X_v}}{G}{c}$.
    We will say that these are \emph{corresponding edges} and this translation is defined as follows.
    For an edge $uc \in E(\ContractXinGtoV{\Compl{X_v}}{G}{c})$ its corresponding edges in $G$ are $\Cut{\{u\}} \cap \Cut{X_v}$ and similarly, if $cc'$ is an edge of $\ContractXinGtoV{\Compl{X_1}}{\ContractXinGtoV{\Compl{X_2}}{G}{c}}{c'}$, for two laminar tight cuts $\Cut{X_1}$ and $\Cut{X_2}$, belonging to two edges of $E(T)$ that share an endpoint, the edges corresponding to $cc'$ are $\Cut{X_1} \cap \Cut{X_2}$.
    We note that this also serves to define corresponding edges for $br(v)$, as $br(v)$ is constructed via tight cut contractions from $G$.
    Further, for an edge set $E_1 \subseteq E(\ContractXinGtoV{\Compl{X_v}}{G}{c})$, we let the corresponding set of edges in $G$ be $(E_1 \cap E(\InducedG{G}{X_v})) \cup \{ uw \in E(G) \mid uc \in E_1 \text{ and } w \in \Compl{X_v} \}$.

    We will also use this correspondence in reverse.
    In particular, if $M$ is some (not necessarily perfect) matching in $G$, then the corresponding matching in some $\ContractXinGtoV{\Compl{X_v}}{G}{c}$ is the unique matching in $\ContractXinGtoV{\Compl{X_v}}{G}{c}$ whose corresponding edge set contains $M$.

    The following objects will be fixed throughout the proof.
    Let $v \in V(T)$ be some vertex, let $G' = \ContractXinGtoV{\Compl{X_v}}{G}{c}$, and let $M'$ be a perfect matching of $G'$.
    Further, let $M$ be a matching of $G$ that corresponds to $M'$ and let $M_v$ be the set of edges of $M$ corresponding to a perfect matching of $br(v)$, which must contain exactly two edges.
    We let $\{ e_v \} = \Cut{X_v} \cap M_v$ and $\{ f_v \} = M_v \setminus \{ e_v \}$.

    We now iteratively construct $F_e(M)$ and $F_f(M) = F_e(M) \setminus \{ e_v \}$ for any (not necessarily perfect) matching $M$ of $G$ that corresponds to a perfect matching of $G'$.
    The difference of these two sets represents the possibility that it may not be optimal to take the edge $e_v$ into the forcing set and instead this edge should be forced later on in the algorithm.
    Thus the focus lies mainly on $F_f(M)$ and we store the size of $F_f(M)$ in the matrix $A$, since the size of $F_e(M)$ is at most one greater.
    (Neither of these sets is explicitly present in the algorithm and they instead get abstracted away into their cost.)

    If $v$ is a leaf, we can define these sets as follows.
    The matching $M$ must consist of an edge $e_v$ from $\Cut{X_v}$ and the single edge $f_v$ that is shared by $G'$ and $G$.
    We let $F_f(M) = \emptyset$ and $F_e(M) = \{ e_v \}$, where it is easy to observe that $F_e$ forces the edge $f_v$ in $G$, since $f_v$ is incident to a vertex of degree two, whose neighbour outside of $f_v$ is contained in $e_v$.
    If we let $e_v'$ be the edge in $G'$ corresponding to $e_v$ and let $F_e'(M) = \{ e_v' \}$, we can again observe that $F_e'$ forces $f_v$ and is thus in particular a smallest forcing set for $M'$ in $G'$.

    Suppose instead that $v$ is an inner node of $T$, possibly even the root.
    Let $k$ and the associated tight cuts $\Cut{Y_i}$ be defined as in line 11 of \Cref{alg:forcing}.
    For every $i \in [k]$, let $M_i \subseteq M$ be the matching that corresponds to a perfect matching of $\ContractXinGtoV{\Compl{Y_i}}{G}{c}$.
    \begin{itemize}
        \item If $\bigcup_{i = 1}^k F_f(M_i) \cup \{ e_v \}$ forces $f_v$ in $G$, we let $F_f(M) = \bigcup_{i = 1}^k F_f(M_i)$.

        \item Otherwise, let $F_f(M) = \bigcup_{i = 1}^k F_f(M_i) \cup \{ f_v \}$.
    \end{itemize}
    We also either let $F_e(M) = F_f(M)$, if $F_f(M)$ forces $e_v$, or we let $F_e(M) = F_f(M) \cup \{ e_v \}$.

    \begin{claim}\label{claim:findforc}
        $F_e(M_0)$ is a smallest forcing set for every $M_0 \in \Perf{G}$.
    \end{claim}
    \emph{Proof of \Cref{claim:findforc}:}
        To prove this claim, we take a smallest forcing set $F$ for $M_0$ and iteratively rearrange it into $F_e(M_0)$, retaining its size and the property that it is a forcing set.
        Let $d = |V(T)|$ and label the vertices of $T$ such that $V(T) = \{ v_1, \ldots , v_d \}$ and for each $v_i$ with the parent $v_j$ we have $i < j$.
        Further, similar to our previous set of fixed objects, for each $i \in [d]$, let $M_{v_i} \subseteq M_0$ be the set consisting of two edges that corresponds to a perfect matching of $br(v_i)$, let $\{ e_{v_i} \} = \Cut{X_{v_i}} \cap M_{v_i}$, and let $M_i'$ be the set of edges in $M_0$ that corresponds to a perfect matching of $\ContractXinGtoV{\Compl{X_{v_i}}}{G}{c}$.
        These objects will only be used in the context of this claim.

        We define sets $F_i$ for all $0 \leq i \leq d$ as follows.
        Let $F_0 = F$ and let $i \in [d]$.
        We set~$F_i = ( F_{i-1} \setminus \{ f_{v_i} \} ) \cup \{ e_{v_i} \}$, if $f_{v_i} \in F_{i-1}$ and $( M_i' \cap F_{i-1} \setminus \{ f_{v_i} \} ) \cup \{ e_{v_i} \}$ forces $f_{v_i}$ in $\ContractXinGtoV{\Compl{X_{v_i}}}{G}{c}$, and otherwise set $F_i = F_{i-1}$.
        
        Our goal is to now prove the following two statements via induction over $[d]$:
        \begin{enumerate}
            \item $( F_i \cap M_i' ) \setminus \{ e_{v_i} \}$ is equal to $F_f(M_i')$ for all $i \in [d]$.\label{item:forcsmallchange}

            \item $F_i$ is a smallest forcing set for $M_0$ in $G$ for all $0 \leq i \leq d$.\label{item:forcsmall}
        \end{enumerate}

        For $i = 0$ \cref{item:forcsmall} holds by definition of $F_0$.
        Let us consider the case $i = 1$, which means that $v_i$ is a leaf of $T$ that is not the root.
        As defined earlier $F_f(M_1')$ does not contain $f_{v_1}$ and our definition for $F_1$ removes $f_{v_1}$ from $F_0$ to construct $F_1$, if it is present there, since either edge of a perfect matching of $C_4$ forces the other.
        Thus \cref{item:forcsmallchange} holds in this case.

        If $F_1 = F_0$, then there is nothing to prove for \cref{item:forcsmall}.
        Suppose instead that $f_{v_1} \in F_0$, which means that $f_{v_1} \not\in F_1$ and $e_{v_1} \in F_1$.
        Since $f_{v_1}$ is adjacent to a vertex of degree two whose non-matched neighbour lies in $e_{v_1}$ it is easy to see that $F_1$ forces $F_0$ and is thus a forcing set for $M_0$.
        In particular, this means that $e_{v_1} \not\in F_0$, as we have otherwise decreased the size of the smallest forcing set.
        Thus $|F_0| = |F_1|$ and \cref{item:forcsmall} holds for $i = 1$.

        Towards proving the induction step for \cref{item:forcsmallchange}, we note that $( F_{i-1} \cap M_j' ) \setminus \{ e_{v_j} \}$ is equal to $F_f(M_j')$ for all $j \in [d]$ such that $v_j \in children(v_i)$, according to the induction hypothesis.
        Therefore
        \[ \bigcup_{v_j \in children(v_i)} F_f(M_j') \cup \{ e_{v_i} \} = ( (M \cap F_{i-1}) \setminus \{ f_{v_i} \} ) \cup \{ e_{v_i} \} . \]
        We now have to show that $f_{v_i} \in F_i$ if and only if $f_{v_i}$ is not forced by $( (M \cap F_{i-1}) \setminus \{ f_{v_i} \} ) \cup \{ e_{v_i} \}$ in $\ContractXinGtoV{\Compl{X_{v_i}}}{G}{c}$, to fit the definition of $F_f(M_i')$ and since $e_{v_i}$ is not contained in either side of the equality we have to prove.
        Suppose that $f_{v_i} \in F_i$.
        Our definition of $F_i$ will never add $f_{v_i}$ and thus $f_{v_i} \in F_{i-1}$.
        Since we have $f_{v_i} \in F_i \cap F_{i-1}$, we conclude that $( (M \cap F_{i-1}) \setminus \{ f_{v_i} \} ) \cup \{ e_{v_i} \}$ does not force $f_{v_i}$ in $\ContractXinGtoV{\Compl{X_{v_i}}}{G}{c}$, confirming the first direction.

        Suppose instead that $f_{v_i} \not\in F_i$.
        If $f_{v_i} \not\in F_{i-1}$, our induction hypothesis tells us that $( (M \cap F_{i-1}) \setminus \{ f_{v_i} \} ) \cup \{ e_{v_i} \}$ forces $f_{v_i}$ in $\ContractXinGtoV{\Compl{X_{v_i}}}{G}{c}$.
        On the other hand, if $f_{v_i} \in F_{i-1}$, then the definition of $F_i$ guarantees that $( (M \cap F_{i-1}) \setminus \{ f_{v_i} \} ) \cup \{ e_{v_i} \}$ forces $f_{v_i}$ in $\ContractXinGtoV{\Compl{X_{v_i}}}{G}{c}$.
        Thus \cref{item:forcsmallchange} holds.

        For the induction step of \cref{item:forcsmall}, if $F_{i-1} = F_i$ there is nothing to prove.
        Thus we may suppose that $F_{i-1}$ and $F_i$ differ.
        By definition $F_i$ contains at most one edge, namely $e_{v_i}$ that $F_{i-1}$ does not contain.
        According to the induction hypothesis $F_{i-1}$ is already a smallest forcing set for $M_0$ in $G$.
        We also know that $F_i$ only contains $e_{v_i}$, if $f_{v_i} \in F_{i-1}$ and $( (M \cap F_{i-1}) \setminus \{ f_{v_i} \} ) \cup \{ e_{v_i} \}$ forces $f_{v_i}$ in $\ContractXinGtoV{\Compl{X_{v_i}}}{G}{c}$.
        Since $( (M \cap F_{i-1}) \setminus \{ f_{v_i} \} ) \cup \{ e_{v_i} \} \subseteq F_i$, we therefore have $|F_i| = |F_{i-1}|$ and as $F_i$ forces $F_{i-1}$, it is itself a forcing set, which proves \cref{item:forcsmall}.

        Using these two statements, we will now show that $F_d = F_e(M_0)$, which immediately implies our claim.
        We note that $M_0 = M_d'$ and thus \cref{item:forcsmallchange} tells us that $F_d \setminus \{ e_r \} = F_f(M_0)$.
        Further, if $e_r \not\in F_d$, then $F_d$ must force $e_r$, according to \cref{item:forcsmall}, and thus $F_d = F_d \setminus \{ e_r \} = F_f(M_0) = F_e(M_0)$ in this case.
        
        We may therefore suppose that $e_r \in F_d$.
        By definition of $F_d$, the set $F_d$ therefore forces $f_r$ and $f_r \not\in F_d$.
        Since $F_d \setminus \{ e_r \} = F_f(M_0)$, we thus know that $f_r \not\in F_f(M_0)$.
        Hence, by definition of $F_f(M_0)$, the set $F_f(M_0) \cup \{ e_r \}$ forces $f_r$ in $G$.
        Consequently, we have $F_e(M_0) = F_f(M_0) \cup \{ e_r \} = F_d$, which proves our claim.
	\hfill$\blacksquare$

    We resume using the objects $v$, $X_v$, $G'$, $M'$, $M$, $M_v$, $e_v$, and $f_v$ that we defined at the beginning of the proof.
    For any $F \subseteq M$, let $Z_v(F)$ be the set of edges in $\Cut{X_v}$ that are not forced out by $F$ in $G$.
    Next we ensure that for $v$ and $M$ the correct tuple $( Z_v(F_f(M)), |F_f(M)| )$ is encoded in $A$, meaning that the corresponding entry of $A$ is set to 1.

    \begin{claim}\label{claim:invariant}
        For any $v \in V(T)$, with a fixed $\Cut{X_v}$ and $e_v \in M$, where $a$ is the lowest index with respect to $<_{X_v}$ in $Z_v(F_f(M))$ and $b$ is the highest such index, we have
        \begin{align*}
            A[\Cut{X_v}, e_v, (y,z), x ] = 1 \Leftrightarrow y = a, \ z = b, \text{ and } x = |F_f(M)| ,
        \end{align*}
        at the end of \Cref{alg:forcing}.
    \end{claim}
    \emph{Proof of \Cref{claim:invariant}:}
        We will prove this claim via induction over $T$, starting from the leaves.
        If $v$ is a leaf, we note that $G' = br(v)$, which is isomorphic to $C_4$.
        Therefore the matchings in the set $S'$ computed in line 14 are all matchings in $G$ that correspond to perfect matchings of $G'$.
        As defined above, for leaves we always set $F_f(M) = \emptyset$ and thus $Z_v(F_f(M)) = \Cut{X_v}$.
        Most of the loop starting in line 15 is not executed in this case, since $v$ has no children.
        Thus we land in the if-statement starting in line 17, which sets $A[\Cut{X_v}, e_v, (1,|\Cut{X_v}|), 0] = 1$, satisfying our claim.

        Suppose instead that $v$ is an interior node or the root of $T$.
        Here we can ignore the lines 17 to 19.
        We let $\Cut{Y_i}$ be defined for all $i \in [k]$ for $k = |children(v)|$ as in line 11 and let $e_1, \ldots , e_k$ be the edges of $M$ contained in the tight cuts $\Cut{Y_1}, \ldots , \Cut{Y_k}$, as in line 20.
        Let $u_i$ be the child of $v$ associated with $\Cut{Y_i}$.
        By our induction hypothesis, the set $T_i$ accurately represents the tuples $((a,b), |F_f(M_i)|)$ for which there exists a matching $M_i$ of $G$ corresponding to a perfect matching of $\ContractXinGtoV{\Compl{Y_i}}{G}{c}$, where $a$ and $b$ represent the lowest and highest index of the edges in $Z_{u_i}(M_i)$ with respect to $<_{Y_i}$.
        The $(a,b)$ suffices to encode $Z_{u_i}(M_i)$, thanks to \Cref{lem:interval}.

        In the for-loop starting in line 22, we consider a combination $(t_1, \ldots , t_k)$ of the tuples from $T_1, \ldots , T_k$.
        Each tuple $t_i = ((a_i,b_i),x)$ defines a set $Z_i$ consisting of the tight cuts edges from $\Cut{Y_i}$ with their indices according to $<_{Y_i}$ lying between $a_i$ and $b_i$, which are not forced out at this point.
        The set $W$ contains all matchings of $S'$ consisting of some combination of edges that are not yet forced out, according to the contents of the tuple $(t_1, \ldots , t_k)$, with the reliability of this information being guaranteed by our induction hypothesis.
        In $W_e$ we then gather those matchings from $W$ that contain $e_v$ and $W_f$ contains the matchings that use $f_v$.
        Since we have $M \in W \cap W_e \cap W_f$, all of these sets are non-empty.

        Should $W_e$ only contain one matching, which must be $M$, then this means that $f_v$ is forced by the partial forcing sets encoded in $(t_1, \ldots , t_k)$.
        Conversely, if $f_v$ is forced by the partial forcing sets encoded in $(t_1, \ldots , t_k)$, we must have $|W_e| = 1$.
        Thus $W_e$ contains exactly one matching if there exist matchings $M_1, \ldots , M_k$ in $G$ such that $f_v$ is forced by the set $\{ e_v \} \cup \bigcup_{i=1}^k F_f(M_i)$ in $G$.
        Should $\{ e_v \} \cup \bigcup_{i=1}^k F_f(M_i)$ force $f_v$, our definition of $F_f(M)$ tells us that $F_f(M) = \bigcup_{i=1}^k F_f(M_i)$.
        Otherwise, $F_f(M) = \{ f_v \} \cup \bigcup_{i=1}^k F_f(M_i)$.
        The lines 28 and 30 then respectively also compute the remaining edges $Z \subseteq \Cut{X_v}$ that are not yet forced out, if $F_f(M)$ does not contain $f_v$ in line 28, and if it does contain $f_v$ in line 30.
        Note that we base line 28 on $W$ and not $W_e$, since we included neither $f_v$ nor $e_v$ in $F_f(M)$.
        
        In either case, line 32 changes the entry $A[\Cut{X_v}, e_v, (a,b), x]$ to 1 exactly for those tuples $((a,b), x)$ where $(a,b)$ encodes $Z$ according to $<_{X_v}$ and $x = |F_f(M)|$, which proves this claim.
    \hfill$\blacksquare$

    It remains to be shown that the set $R$ that is returned by \Cref{alg:forcing} is in fact $\FSpec{G}$.
    We note that if $v = r$, the graph $G'$ is isomorphic to $G$, due to the choice of $X_r$ in lines 6--8.
    First, we must show that $\Forc{G,M}$ is actually contained in $R$.
    Suppose $F$ is a smallest forcing set for $M$, then \Cref{claim:findforc} tells us that $F_e(M)$ is a smallest forcing set for $M$ and thus has the same size as $F$.
    \Cref{claim:invariant} then tells us that $A[\Cut{X_r}, e, (a,b), |F_f(M)|] = 1$ holds, with $\{ e \} = M \cap \Cut{X_r}$ and $(a,b)$ being the pair of indices describing the edges of $\Cut{X_r}$ that are not forced out by $F_f(M)$.
    
    By construction, $|F_e(M)| \in \{ |F_f(M)|, |F_f(M)| + 1 \}$.
    If $|F_f(M)| \neq |F_e(M)|$, then by definition of $F_e(M)$ this must mean that $e$ is in fact not forced by $F_f(M)$ and thus $a \neq b$, which means that $\mathbbm{1}(a,b) = 0$ and therefore $|F_f(M)| + 1 - \mathbbm{1}(a,b) = |F_f(M)| + 1 = |F_e(M)|$.
    Thus $|F_e(M)| \in R$.
    Otherwise $|F_f(M)| = |F_e(M)|$, $a = b$, and thus $|F_f(M)| + 1 - \mathbbm{1}(a,b) = |F_f(M)| = |F_e(M)|$.
    Therefore we again have $|F_e(M)| \in R$.
    Hence the forcing number of $M$ is found in $R$.

    For the other direction, we need to show that for every value $x$ in $R$ there exists a perfect matching of $G$ that has the forcing number $x$.
    This can easily be seen via the fact that for each $x \in R$ we have $x = |F_e(M)| + 1 - \mathbbm{1}(a,b)$ for some tuple $((a,b), |F_e(M)|)$ such that $A[\Cut{X_r}, e, (a,b), |F_e(M)|] = 1$, where $(a,b)$ encodes $Z_r(F_f(M))$.
    According to \Cref{claim:invariant} this means that $M$ is a perfect matching of $\ContractXinGtoV{\Compl{X_r}}{G}{w}$ containing $e$.
    Since $\ContractXinGtoV{\Compl{X_r}}{G}{w}$ and $G$ are isomorphic, we can treat $M$ as a perfect matching of $G$ and $F_e(M)$ as a forcing set for $M$, which according to \Cref{claim:findforc} is a smallest forcing set for $M$ in $G$.
    Thus $x$ accurately reflects the size of a smallest forcing set in $G$.
    We have therefore proven that $R = \FSpec{G}$, which concludes the proof of our theorem.
\end{proof}

Analysing the runtime of the algorithm is much less of a hassle.

\begin{theorem}\label{thm:runtimealgo}
    Given a matching covered, outerplanar graph $G$ as the input, \Cref{alg:forcing} computes $\FSpec{G}$ in $\mathcal{O}(n^{15})$-time.
\end{theorem}
\begin{proof}
    We start by noting that for $G$ we have $m \in \mathcal{O}(n)$, where $n = |V(G)|$ and $m = |E(G)|$, which can for example be seen by using Euler's formula.
    To compute the tight cut decomposition in line 2, we can use the method from \Cref{lem:onlyc4braces}.
    Finding a vertex of degree exactly 2 can be done in $\mathcal{O}(n)$-time and this procedure has to be repeated at most $\mathcal{O}(n)$ times, landing us in $\mathcal{O}(n^2)$-time for the computation of the tight cut decomposition $(T, \beta, \pi)$, where we note that $|V(T)| \in \mathcal{O}(n)$.
    Thus all operations in lines 3--8 run in $\mathcal{O}(n^2)$-time and setting up the matrix $A$ can be done in $\mathcal{O}(n^5)$-time.

    The while-loop starting in line 10 has $\mathcal{O}(n)$ iterations and thanks to \Cref{lem:onlyc4braces}, we know that for all $v \in V(T)$ the graph $br(v)$ has a constant size and $v$ has at most four neighbours.
    This means that the lines 11--14 take at worst $\mathcal{O}(n^2)$-time each iteration and $S'$ contains $\mathcal{O}(n^2)$ elements, which each correspond to an iteration in the for-loop starting in line 15.
    The lines 16--20 are again not costly, running in linear time for each iteration of the for-loop.
    Line 21 however needs to check $\mathcal{O}(n^3)$ tuples for each $i \in [k]$, with $k \leq 3$.
    In the worst case, this leaves us with $\mathcal{O}(n^9)$ tuples which gives us the number of iterations of the for-loop starting in line 22.

    The set $W$ has a size in $\mathcal{O}(n^2)$, since it is a subset of $S'$, and it takes $\mathcal{O}(n)$ time to check each of their intersections with the sets $Z_1, \ldots , Z_k$, of which there are at most three, leaving us with a runtime in $\mathcal{O}(n^3)$ for line 24.
    In the lines 25--33 everything is again computable in $\mathcal{O}(n^2)$-time.
    To put together our solution set in line 34, we have to scan a small part of our table.
    However the first entry is fixed and for the second and third entry we know that there are a constant set of possible values, leaving us at $\mathcal{O}(n)$-time for this step.

    Altogether our nested loops thus run in $\mathcal{O}(n^{15})$-time.
\end{proof}

All that is left now is to prove the main theorem.
The main hurdle to its proof was \Cref{thm:correctnessalgo} and the runtime is simply bounded from above by the runtime given in \Cref{thm:runtimealgo}.

\begin{proof}[Proof of \Cref{thm:main}]
    Given an outerplanar graph $G$ there is nothing to do unless $G$ has a perfect matching, which we will therefore assume exists.
    Since $G$ is not necessarily bipartite, we can use the algorithm in \cite{CarvalhoC2005VE} to compute its cover graph, yielding some connected components $G_1, \ldots , G_k$ for some integer $k < n$.
    For each of these components, we can then apply \Cref{alg:forcing}, which returns their forcing spectrum according to \Cref{thm:correctnessalgo} and runs in $\mathcal{O}(n^{15})$-time according to \Cref{thm:runtimealgo}.
    It can then easily be verified that
    \[ \FSpec{G} = \{ \sum_{i=1}^k x_i \mid x_i \in \FSpec{G_i} \} , \]
    which can be computed in $\mathcal{O}(n^3)$-time.
    This leads to a total runtime in $\mathcal{O}(n^{15})$ and thus the theorem is proven.
\end{proof}

%% file: discussion.tex

Given the apparent simplicity of the class of outerplanar graphs, the exponent in the runtime of our algorithm is quite large.
Though we hope that faster algorithms for computing the forcing number of outerplanar graphs can be found, we believe that the problem itself is inherently so complex that the size of our exponent is not unsurprising.

The main contributor to the runtime is the process in line 21 of \Cref{alg:forcing}, which could possibly be replaced with a more clever data-structure.
Other than this, we believe that since our algorithm inherently orients itself on methods for FPT-algorithms, this method likely will not directly yield low runtimes.
Instead we hope that our approach to the problem via structural matching theory leads us closer to a structure in outerplanar graphs that may be exploited in a more elegant way.

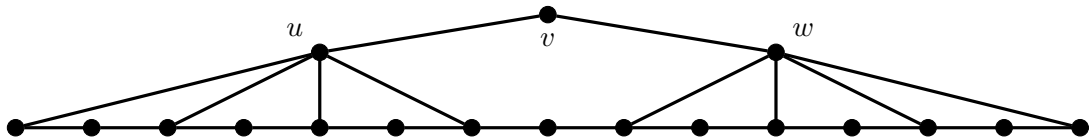
\begin{figure}[h!]
	\begin{center}
		\input{graphs/maxdeg}
	\end{center}
	\caption{A matching covered, outerplanar graph with a tight cut around $\{ u,v,w \}$.}
	\label{fig:maxdeg}
\end{figure}

In this context, we want to point out that \Cref{lem:interval} cannot be replaced with a bound on the maximum degree of the graph or an argument about the size of tight cuts in outerplanar graphs, since neither is bounded by a constant in matching covered, outerplanar graphs, as can be seen by extrapolating from the graph we give in \Cref{fig:maxdeg}.

\subsection{Potential generalisations of the algorithm}
Naturally one might wonder if this algorithm can be generalised to broader classes of graphs.
Two come to mind immediately, with the first being those matching covered graphs who have no bricks and only have $C_4$ as their brace.
Sadly it is not clear how to limit ourselves to a polynomial number of open edges to check in the tight cuts throughout the algorithm.
The structural approach to controlling the number of these sets in \Cref{sec:outerplanar} relies not only on the Hamiltonicity of our graphs, but also requires the outerplanar drawing.
Though we note that this class still has nice structural properties and for example corresponds to the graphs with directed treewidth one \cite{Wiederrecht2020Digraphs}.

The second class one might want to tackle is planar graphs, in hopes of resolving \Cref{que:forcplanar}.
Here the problem is that it is easy to find planar graphs which cannot be decomposed into small bricks and braces.
In particular one can construct arbitrarily large cubic, planar braces (see \cite{GorskySW2023Matching}).
Thus our approach is not particularly helpful in this setting.

In general, for a generalisation of our algorithm one needs the following ingredients:
\begin{itemize}
    \item A graph class $\CCC$ that consists of matching covered graphs whose bricks and braces belong to a finite set of graphs, or alternatively to a set of graphs with only polynomially many perfect matchings in terms of the size of any given graph in the set.

    \item The tight cuts of graphs in $\CCC$ need to be controllable enough to allow us to check at most polynomially many sets of potentially open edges in terms of the size of the graph.

    \item The tree $T$ in tight cut decomposition $(T, \beta, \pi)$ of any graph in $\CCC$ must have its maximum degree bounded by some constant.
    This is automatically true if the bricks and braces come from a fixed finite set of graphs.
\end{itemize}

\subsection{Trying to compute the anti-forcing number in outerplanar graphs}
The anti-forcing number is a natural companion of the forcing number, which focuses on deleting edges outside of the perfect matching we try to force.
This parameter was introduced in \cite{VukieevicT2007Antiforcing} and first implicitly used in \cite{LukovitsMTV2005Kekule} (see also \cite{HararyKZ1991Graphical,Li1997Hexagonal}).

\begin{definition}[The anti-forcing number]\label{def:antiforce}
    Given a graph $G$ and some $M \in \Perf{G}$, an \emph{anti-forcing set $F \subseteq E(G) \setminus M$ for $M$} is a set of edges such that the unique perfect matching of $G - F$ is $M$.
    The \emph{anti-forcing number of $M$ in $G$}, denoted as $\AForc{G,M}$ is the minimum size of an anti-forcing set $F \subseteq E(G) \setminus M$ for $M$\footnote{Though this is a natural variant of the forcing number $\Forc{G,M}$, the first mention of this specific parameter is found in \cite{LeiYZ2016Antiforcing}.} and the \emph{anti-forcing-spectrum} $\AFSpec{G}$ is then simply defined as $\{ \AForc{G,M} \mid M \in \Perf{G} \}$.

    The \emph{minimum anti-forcing number} $\AForc{G}$ is simply $\min(\AFSpec{G})$. 
\end{definition}

Similar to the forcing number, it is known that computing $\AForc{G,M}$ is $\NP$-complete if $G$ is a bipartite graph of maximum degree 4 \cite{DengZ2017Antiforcinga} and if $G$ is bipartite and planar, one can again translate $G$ into a digraph and use the result by Gabow \cite{Gabow1995Centroids} to determine $\AForc{G,M}$ in polynomial time.
Determining the computational complexity of finding $\AForc{G}$ is an open question even in planar bipartite graphs \cite{DengZ2017Antiforcinga} and the analogous formulation of \Cref{que:forcnonbipplanar} is also relevant for this parameter.

\begin{question}\label{que:antiforcnonbipplanar}
    What is the complexity of determining $\AForc{G,M}$ for a non-bipartite, planar graph $G$ with a perfect matching $M$?
\end{question}

If one were to try to adapt our algorithm for the purpose of computing the anti-forcing number of outerplanar graphs, then, via entirely analogous methods, it is easy to reproduce the contents of \Cref{sec:outerplanar} for the anti-forcing number.

\begin{lemma}
     If $G$ is a graph with a perfect matching $M$, then $\AForc{G,M} = \AForc{\Cov{G},M}$ and $\AFSpec{G} = \AFSpec{\Cov{G}}$.
\end{lemma}

\begin{lemma}
    Let $G$ be a matching covered, outerplanar graph with a tight cut $\Cut{X}$ and a perfect matching $M$.
    Further, let $F \subseteq E(\InducedG{G}{X}) \setminus M$.
    The number of different sets of edges from $\Cut{X}$ that are not forced out by $F$ is polynomial in $|\Cut{X}|$. 
\end{lemma}

However the troubles lies in controlling how many different partial anti-forcing sets need to be considered for each vertex $v$ of the tree of the tight cut decomposition in \Cref{alg:forcing}.
When considering the forcing number, it is clear that we only need to delete some subset of the matching corresponding to a perfect matching of $br(v)$ to test our partial forcing sets.
Since each brace here is $C_4$, there is only a constant number of perfect matchings of $br(v)$ and at worst $\mathcal{O}(n^2)$ many matchings corresponding to these matchings.
However the number of potential anti-forcing sets corresponding to edges of $br(v)$ outside of the matching is at worst in $\mathcal{O}(2^n)$.
This seems like a major hurdle that requires some new tricks.

%% file: graphs/maxdeg.tex
			\begin{tikzpicture}

            \foreach\i in {1,...,15}
                {
                    \node (V\i) at (\i,0) [draw, circle, scale=0.6, fill] {};
                }
            \node (U1) at (5,1) [draw, circle, scale=0.6, fill, label=north west:{$u$}] {};
            \node (U2) at (8,1.5) [draw, circle, scale=0.6, fill, label=south:{$v$}] {};
            \node (U3) at (11,1) [draw, circle, scale=0.6, fill, label=north east:{$w$}] {};

            \foreach\i in {1,...,14}
                {
                    \pgfmathsetmacro\iplus{\i+1}
                    \path (V\i) edge[very thick] (V\iplus);
                }
            \foreach\i in {1,3,5,7}
                {
                    \path (V\i) edge[very thick] (U1);
                }
            \foreach\i in {9,11,13,15}
                {
                    \path (V\i) edge[very thick] (U3);
                }

            \path
                (U1) edge[very thick] (U2)
                (U2) edge[very thick] (U3)
            ;
			\end{tikzpicture}